\documentclass[preprint,showpacs,amsmath,amssymb,preprintnumbers]{revtex4}
\usepackage{graphicx}
\usepackage{amsmath}

\begin{document}

\title{Modal interactions of flexural and torsional vibrations in a microcantilever}
\author{H. J. R. Westra} \email{h.j.r.westra@tudelft.nl}
\author{H. S. J. van der Zant}
\author{W. J. Venstra} 
\affiliation{Kavli Institute of Nanoscience, Delft University of Technology, Lorentzweg 1, 2628CJ Delft, The Netherlands}

\date{\today}
\begin{abstract}
The nonlinear interactions between flexural and torsional modes of a microcantilever are experimentally studied. The coupling is demonstrated by measuring the frequency response of one mode, which is sensitive to the motion of another resonance mode. The flexural-flexural, torsional-torsional and flexural-torsional modes are coupled due to nonlinearities, which affect the dynamics at high vibration amplitudes and cause the resonance frequency of one mode to depend on the amplitude of the other modes. We also investigate the nonlinear dynamics of torsional modes, which cause a frequency stiffening of the response. By simultaneously driving another torsional mode in the nonlinear regime, the nonlinear response is tuned from stiffening to weakening. By balancing the positive and negative cubic nonlinearities a linear response is obtained for the strongly driven system. The nonlinear modal interactions play an important role in the dynamics of multi-mode scanning probe microscopes.
\end{abstract}

\maketitle

\section{Introduction}
\indent \indent The Atomic Force Microscope (AFM)~\cite{Binnig1986:p930} is a crucial instrument in studying nanoscale objects. Various operation schemes are employed, which include the use of different cantilever geometries, higher modes or the torsional mode for imaging~\cite{Huang2004277,Neumeister1994:p2527,Pires2010:p0705,Dong2009}. The nonlinear tip-sample interactions determine the dynamics in tapping-mode AFM and have been studied in detail~\cite{hillenbrand2000p3478,Rutzel08082003}. Besides this extrinsic nonlinearity, the intrinsic mechanical nonlinearities determine the dynamics of ultra-flexible microcantilevers at high amplitudes, as shown in a recent study~\cite{Venstra:2010p7325}. These nonlinearities result in a amplitude-dependent resonance frequency and couple the vibration modes. In clamped-clamped beams, the nonlinear coupling is provided by the displacement-induced tension~\cite{Westra:2010p7074,Gaidarzhy2011:p264106}. For cantilever beams it was shown that the coupling between the modes can be used to modify the resonance linewidth~\cite{Venstra2011:p151904}. In a multi-mode AFM~\cite{Garcia2012,Raman2011:p809}, these modal interactions are of importance, since the resonance frequency of one mode depends on the amplitude of the other modes.\\
\indent \indent In this work, we experimentally demonstrate the intrinsic mechanical coupling between the flexural and torsional modes of a microcantilever. The resonance frequency of one mode depends on the amplitude of the other modes. The flexural modes are coupled via the geometric and inertial nonlinearities. The torsional modes exhibit frequency stiffening at high amplitudes, which originates from torsion warping~\cite{Sapountzakis2010:p1853}. Interestingly, the nonlinearity constant of one torsional mode changes sign when another torsional mode is driven at high amplitudes. Finally, the coupling between the torsional and flexural modes is studied.\\

\section{Experiment}
\indent \indent Microcantilevers are fabricated by photolithographic patterning of a thin low-pressure chemical vapor deposited silicon nitride (SiN) film. Subsequent reactive ion etching transfers the pattern to the SiN layer, and the cantilevers are released using a wet potassium hydroxide etch, resulting in a undercut-free cantilever. The dimensions are length $\times$ width $\times$ height ($L \times w \times h) = 42 \times 8 \times 0.07$ $\mu\mathrm{m}^3$. These floppy cantilevers allow high amplitudes and thus facilitate the study of nonlinearities. The cantilever is mounted onto a piezo actuator, which is used to excite the cantilever. The cantilevers are placed in vacuum (pressure $< 10^{-5}$ mbar) to eliminate air-damping and to enable large vibration amplitudes, where nonlinear terms in the equation of motion dominate the dynamics. The cantilever motion is detected using a home-made optical deflection setup which resembles the detection scheme frequently used in scanning probe microscopes. The flexural and torsional vibration modes are detected with a sensitivity of $\pm$ 1 pm/$\sqrt{\mathrm{Hz}}$~\cite{BabaeiGavan:2009p4633}. A schematic of the measurement setup is shown in Fig. 1(a). The cantilever displacement signal is measured using either a network (NA) or spectrum analyzer (SA). To drive a second mode, a separate RF source is used.\\

\begin{figure}[!h!t]
\includegraphics[width=135mm]{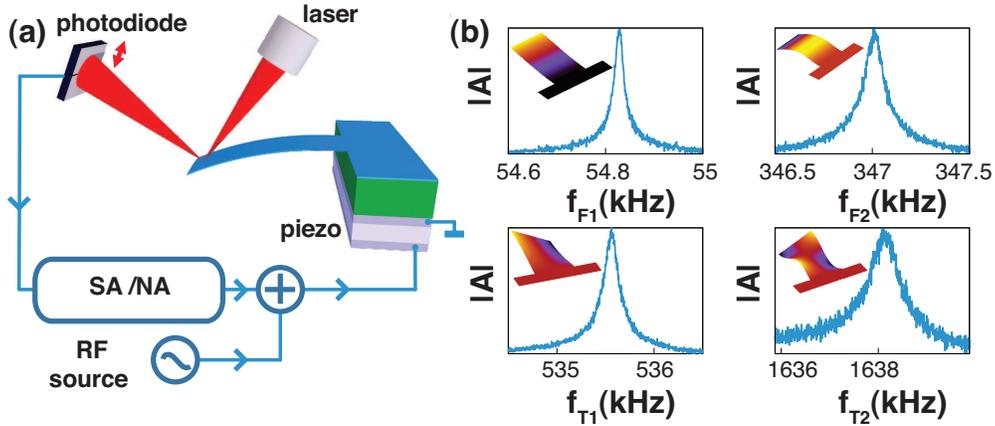}
\caption{Measurement setup. (a) Optical deflection setup showing the laser beam, which reflects from the cantilever surface. The spot of the reflected laser beam is modulated in time by a frequency corresponding to the cantilever motion. The cantilever is mounted onto a piezo actuator in vacuum. Network (NA) and spectrum analysis (SA) is performed on the signal from the two-segment photodiode. (b) Frequency responses of the first and second flexural (top panels) and torsional (bottom panels) modes. Inset are the calculated mode shapes from Euler-Bernoulli beam theory.}
\label{fig1}
\end{figure}

\indent \indent First, the flexural vibrations are characterized by measuring the cantilever frequency response at different resonance modes. The first flexural mode shown in Fig. 1(b) occurs at 54.8 kHz with a Q-factor of 3000. The resonance frequency of the second mode is 347 kHz ($Q=3900$), which is 6.33 times higher than the first resonance mode, in agreement with the calculated ratio $f_{R,F2} / f_{R,F1} = 6.27$, following from Euler-Bernoulli beam theory. Not shown is the third flexural mode at 974.9 kHz, with $f_{R,F3} / f_{R,F1} = 17.8$, near the expected ratio of 17.6. This indicates that in the linear regime the cantilever beam is described by the Euler-Bernoulli beam theory. Throughout the manuscript, the subscripts $\mathrm{F}i$ and $\mathrm{T}i$ indicate the frequency span around the $i^\mathrm{th}$ flexural ($\mathrm{F}$) or torsional ($\mathrm{T}$) resonance mode. The subscript $\mathrm{R}$ refers to the resonance frequency of that particular mode. \\ 
\indent \indent The torsional modes are characterized by rotating the cantilever over 90 degrees in the setup; the two-segment photodiode is then sensitive to vibrations corresponding to torsional resonance modes~\cite{BabaeiGavan:2009p4633}. The frequency response of the first two torsional modes is shown in Fig. 1(b). From theory, the ratio between the lowest two resonance frequencies of the torsional modes is 3, which is close to the measured ratio of $f_{R,T2} / f_{R,T1} = 1638\, \mathrm{kHz} / 535.4\, \mathrm{kHz} = 3.06$. The Q-factors of the first and second torsional mode are 4300 and 3200 respectively.\\
\indent \indent At high drive amplitudes, the flexural and torsional modes become nonlinear. The nonlinearity of the flexural modes in a cantilever beam was theoretically studied by Crespo da Silva in 1978~\cite{CrespodaSilva:1978p30, CrespodaSilva:1978p29}. To include the torsional nonlinearity, the equations of motion are extended (Appendix A). For the flexural and torsional modes, the nonlinearity causes a Duffing-like frequnecy stiffening when the mode is strongly driven~\cite{Lifshitz:2008p7422,Nayfeh:1995p42} leading to a bistable vibration amplitude. This bifurcation is observed in all modes studied in this paper. These nonlinearities are responsible for the coupling between the flexural-flexural, torsional-torsional and flexural-torsional modes.

\section{Modal interactions in a microcantilever}
We now experimentally demonstrate the coupling between the modes of a microcantilever. We use a two-frequency drive signal to excite two resonance modes of the cantilever simultaneously while we measure the motion of one mode. First, we focus on the interactions between the flexural modes. Then we turn our attention to the torsional modes, starting with the amplitude-dependent resonance frequency of the torsional vibrations, followed by the demonstration of the coupling between the lowest two torsional modes. Finally, the interactions between flexural and torsional modes are discussed.

\subsection{Flexural-flexural mode interaction}
\begin{center}
\begin{figure}[!h!t]
\includegraphics[width=135mm]{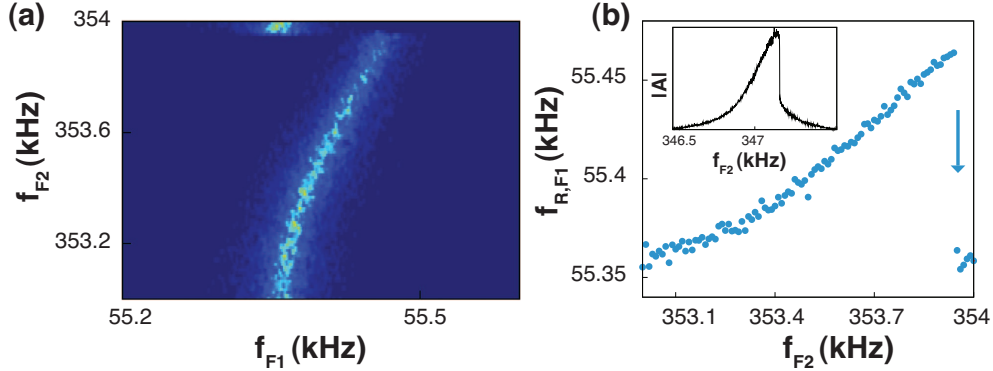}
\caption{Flexural-flexural mode interactions. (a) Frequency spectra of the thermal motion of the first flexural mode ($f_{\mathrm{F1}}$), when the second mode is driven through its resonance frequency. Colorscale represents the power spectral density of the displacement noise of mode 1. As the peak width remains constant, there is no significant change in the Q-factor. The motion of the second mode tunes the resonance frequency of the first mode. (b) The resonance frequency of the first mode $f_\mathrm{R,F1}$ versus the drive frequency of the second flexural mode. The nonlinear response of the second mode is reflected in the fitted resonance frequency of the first mode. Inset: the direct measurement of the nonlinear frequency response of the second mode~\cite{UMfootnote} }
\label{fig2}
\end{figure}
\end{center}

To investigate the interactions between the two lowest flexural modes, the thermal motion of the first mode is measured with a spectrum analyzer, while the RF source strongly drives the second mode. The thermal noise spectra of the first mode as a function of the drive frequency of the second mode are shown in Fig.~\ref{fig2}(a). The color scale represents the power spectral density of the displacement around the resonance frequency of the first mode. A shift of the resonance peak of the first mode is observed as the drive signal at $f_{\mathrm{F2}}$ approaches the nonlinear resonance of the second mode. The resonance frequency of the first mode for each drive frequency of the second mode is obtained by fitting the damped-driven harmonic oscillator (DDHO) response. In Fig.~\ref{fig2}(b), this resonance frequency of the first mode is plotted versus the drive frequency of the second mode. The nonlinear response of the second mode is reflected in the resonance frequency of the first mode where the resonance frequency first increases and then jumps down after the second mode has reached its maximum amplitude, indicated by the arrow. At the maximum amplitude of the second mode, the resonance frequency of the first mode is shifted by several times its linewidth. This experiment shows that the coupling between the flexural modes can introduce significant resonance frequency shifts when multiple modes are excited simultaneously. Moreover, by measuring the shift in resonance frequency of the first mode the motion of the second mode can be detected. For comparison, the nonlinear response of the direct-driven second mode is shown in the inset of Fig.~\ref{fig2}(b).\\

\subsection{Torsional-torsional mode interaction}
\begin{figure}[!h!t]
\includegraphics[width=135mm]{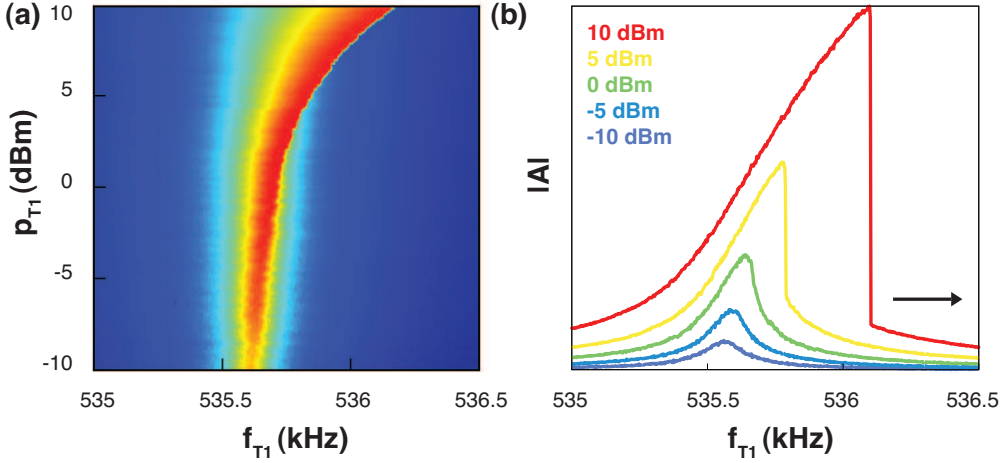}
\caption{Nonlinear torsional mode. (a) Frequency responses of the first torsional mode, when the drive amplitude is increased. Beyond a power of 5 dBm, the response is bistable. Color scale indicates the amplitude normalized to the drive voltage. (b) Resonator amplitude traces at 5 selected drive powers. The nonlinear frequency response is visible at high drive powers. The frequency is swept from low to high. }
\label{fig3a}
\end{figure}

Before turning to the interactions between the torsional vibration modes, we first measure the frequency response of a single torsional mode as a function of the drive strength. Although torsional modes are extensively used in AFMs~\cite{Huang2004277,Lohndorf2000p1176}, their nonlinear behavior has not been investigated in detail. To investigate the nonlinearity, we strongly drive the torsional mode. In contrast to the flexural-flexural modes, where the nonlinearity arises from geometric and inertial effects, in torsional modes, the nonlinearity originates from torsion warping and inertial moments~\cite{Sapountzakis2010:p1853}. In Appendix A we discuss the equations of motion including the nonlinearities involved. The amplitude of the first torsional mode with varying drive power is shown in Fig.~\ref{fig3a}(a), with selected frequency responses of the first torsional mode plotted in Fig.~\ref{fig3a}(b). At low driving power, the resonance line shape is a DDHO response, and the cantilever is oscillating in the linear regime. When the power is increased, the frequency response leans towards higher frequencies and the amplitude bifurcates. Close to this critical amplitude (0 dbm) the slope of the frequency response approaches infinity, which may be used to enhance the sensitivity in torsional mode AFM. A frequency stiffening is observed for the first and second torsional mode.\\

\begin{figure}[!h!t]
\includegraphics[width=135mm]{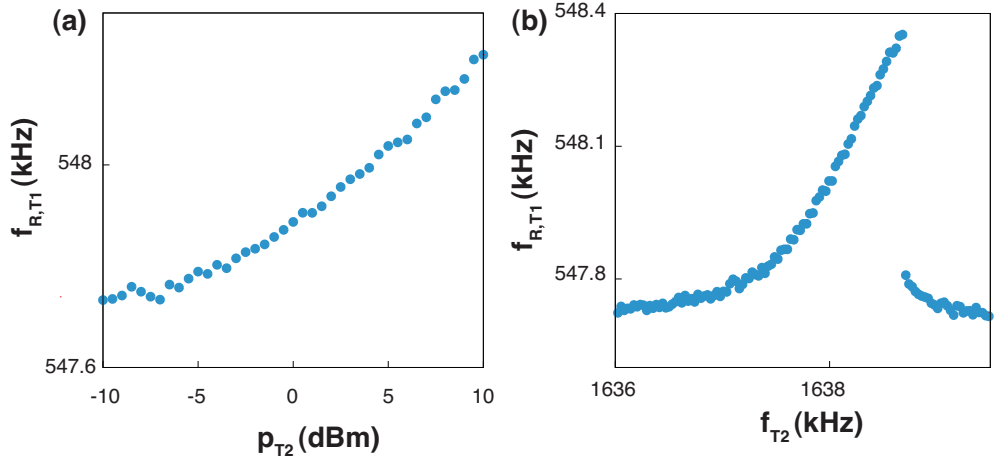}
\caption{Torsional-torsional mode interactions. (a) The resonance frequency of the first torsional mode, when the drive power at the second torsional mode on resonance is varied. The resonance frequency of mode 1, $f_{\mathrm{R,T1}}$, increases with the drive power of the second mode,  $p_{\mathrm{T2}}$. (b) The nonlinear response of the second mode is measured by using the first mode as a detector.}
\label{fig3b}
\end{figure}

\indent \indent Similar to the flexural-flexural modal interactions discussed in the previous section, the coupling between the first and second torsional modes is studied: we measure the thermal noise of the first mode while the drive power of the second torsional mode is varied. The resonance frequencies, obtained from DDHO fits to the thermal noise spectra of the first mode, are shown in Fig.~\ref{fig3b}(a). The resonance frequency increases with 500 Hz, while increasing the driving strength of the second mode to 10 dBm. We now perform a similar experiment as the one as shown in Fig.~\ref{fig2}(b). Thus, the first torsional mode is used to detect the nonlinear vibrations of the second torsional mode. Fig.~\ref{fig3b}(b)) shows the nonlinear response, resembling the behavior of the first mode shown in Fig.~\ref{fig3a}(b).\\

\begin{center}
\begin{figure}[!h!t]
\includegraphics[width=135mm]{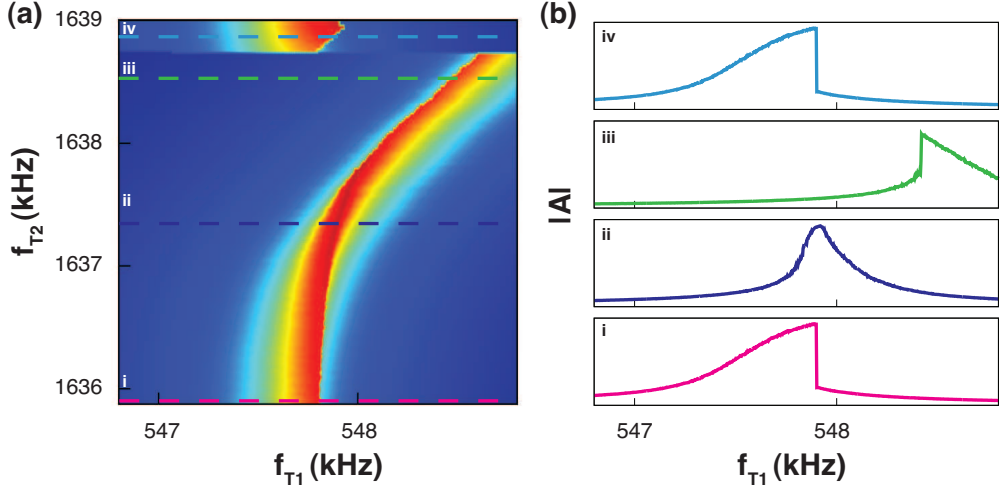}
\caption{Tuning the torsional nonlinearity via modal interactions. (a) Frequency responses of the nonlinear first torsional mode, while the frequency of the second mode is swept trough its nonlinear resonance. The frequency stiffening of the torsional mode changes into weakening when the second mode oscillates with high amplitudes. When the amplitude of the second mode jumps down, again frequency stiffening is observed. (b) Traces from (a) taken at the indicated frequencies. In panel (ii) the response is close to a linear one, due to balancing of the stiffening and weakening nonlinearities.}
\label{fig4}
\end{figure}
\end{center}

\indent \indent Interesting behavior is observed when both torsional modes are driven in the nonlinear regime. In contrast to measurements in the previous section, the first mode is now also driven in the nonlinear regime. Fig.~\ref{fig4}(a) shows the nonlinear frequency response of the first torsional mode, while stepping the drive frequency of the second torsional mode through its resonance. Fig.~\ref{fig4}(b) shows individual traces, which reveal interesting behavior; in the lowest panel (i), there is no influence of the second mode and frequency stiffening of the first mode is observed cf. Fig.~\ref{fig3a}(b). When the amplitude of the second mode starts to increase as it approaches its resonance, the response of mode 1 becomes more linear (panel ii). Here, the frequency stiffening and weakening nonlinearities are balanced yielding a linear response. At high amplitude of the second mode, frequency weakening of mode 1 (panel iii) is observed. When the amplitude of the second mode drops, frequency stiffening is restored (panel iv). This measurement not only demonstrates the coupling between the torsional modes, but also that the sign of the nonlinearity constant of a torsional mode depends on the amplitude of the motion of the other modes. By simultaneous driving another mode, the torsional frequency response can be tuned from a stiffening to a weakening characteristic. \\

\subsection{Flexural-torsional mode interaction}
\begin{center}
\begin{figure}[!h!t]
\includegraphics[width=135mm]{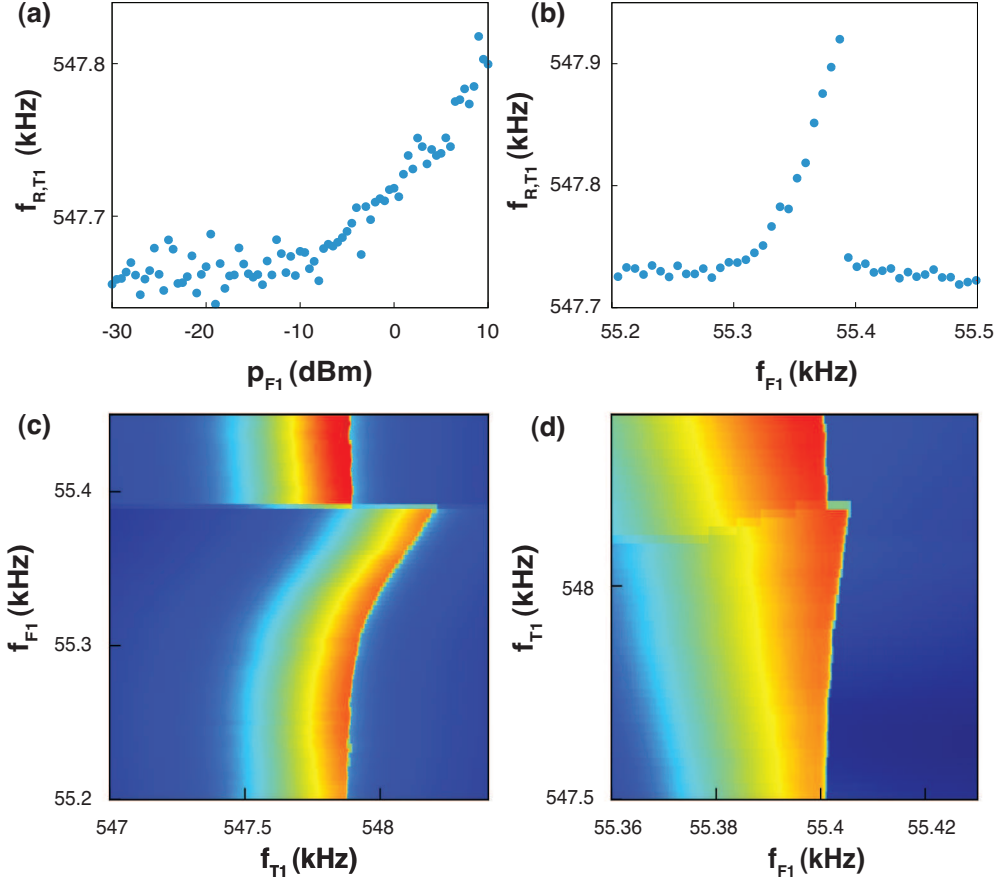}
\caption{Flexural-torsional interaction. (a) Resonance frequency shift of the first torsional mode, when the drive power of the first flexural mode is increased. The resonance frequencies are obtained from thermal noise spectra. (b) The nonlinear resonance response of the first flexural mode reflected in the resonance frequency of the torsional mode (from thermal noise spectra). (c) and (d) Nonlinear dynamics when the flexural as the torsional mode are both excited in the nonlinear regime.}
\label{fig5}
\end{figure}
\end{center}

The coupling between the first flexural and first torsional mode is now studied experimentally. This coupling is theoretically described in Eq.~\ref{eq2}. Fig.~\ref{fig5}(a) shows the resonance frequency of the first torsional mode as a function of drive power of the first flexural mode. The resonance frequency increases with 100 Hz when the power in the flexural mode is increased to 10 dBm. Detection of the nonlinear flexural mode by measuring the resonance frequency of the torsional mode is shown in Fig.~\ref{fig5}(b). The nonlinear interactions when both modes are driven in the nonlinear regime are shown in Fig.~\ref{fig5}(c) and (d). The interaction is clearly visible in the frequency-frequency plots, where one frequency is swept and the frequency of the RF source is stepped across the nonlinear resonances of both modes.\\

\section{Discussion and conclusion} 
In summary, we demonstrated the coupling between the flexural and torsional vibration modes in a microcantilever. This coupling is due to nonlinearities, which also give rise to a amplitude-dependent resonance frequency. The interactions between the different flexural modes, between different torsional modes and between the flexural and torsional modes are demonstrated in detailed experiments. We also demonstrate the nonlinear frequency stiffening of torsional modes driven at high amplitudes. \\ 
\indent \indent Several applications are proposed for the modal interactions. A specific resonance mode can be shifted to a higher frequency by simultaneously driving another mode. For strongly driven modes, the cubic spring constant (nonlinearity) can be modified from positive to negative, tuning the response from stiffening to weakening. By balancing two excitation strengths, a nonlinear response can be tuned to a linear one. By modal interactions, one mode can be used to detect the motion of another mode of the same cantilever. Besides these applications, the modal interactions have consequences for multi-mode schemes, such as the scanning probe microscopy and mass sensors based on microcantilevers~\cite{Raman2011:p809,Li:2007p25, Raiteri2001115,dohn2007p103303}. 

\section*{Acknowledgement}
The authors acknowledge financial support from the Dutch funding organization FOM (Program 10, Physics for Technology).

\newpage

\appendix 
\section{Theory of modal interactions} \label{app}
In this appendix the nonlinear equations of motion of the modes in a cantilever are described. We start with the general equations of motion, which include the coupling between the torsional and flexural modes. Then, we consider flexural modes along one axis. We conclude with the equation of motion of two coupled flexural modes, relevant to the experiment in Section 3.1. We start with the equations derived by Crespo da Silva~\cite{CrespodaSilva:1978p30, CrespodaSilva:1978p29}, who described the motion in two flexural directions $v$ and $w$ and the torsional angle $\theta$:
\begin{align}
m v_{tt} &+ \gamma^v v_t + \frac{\partial^2}{\partial s^2}(D_{\zeta}v_{ss}) = \frac{\partial }{\partial s}\Bigg\{ -D_{\xi} w_{ss}(\theta_s + v_{ss}w_s) + w_s Q_{\theta} - v_s\frac{\partial}{\partial s}\Big[D_{\zeta}(v_s v_{ss} + w_{s}w_{ss})\Big] \nonumber \\
 & + (D_\eta - D_\zeta)w_s v_{ss} w_{ss}  + \Big[ \Big(D_\eta - D_\zeta\Big)\frac{\partial}{\partial s} \Big(\theta w_{ss} - \theta^2 v_{ss}\Big)\Big]  \nonumber \\
 &+ \jmath_\xi w_{st}(\theta_t + v_{st}w_s) + \jmath_\zeta v_{st}\frac{\partial}{\partial t}\Big(v_{s}v_{st} + w_{st}w_{s}\Big) - \frac{\partial}{\partial t}\Big[(\jmath_\eta - \jmath_\zeta)(\theta w_{st} \nonumber \\
 &- \theta^2 v_{st} + w_s v_{st} w_{st}) - \jmath_\zeta v_{st}\Big] - \frac{v_s}{2}\int_L^s \mathrm{d}s' \, m \int_0^{s'} \mathrm{d}s'' \, \frac{\partial^2}{\partial t^2} \Big(v_s^2 + w_s^2\Big) \Bigg\} + q \cos(\Omega t) , \\
m w_{tt} &+ \gamma^q w_t + \frac{\partial^2}{\partial s^2}(D_{\eta}w_{ss}) = \frac{\partial }{\partial s} \Bigg\{ D_{\xi} v_{ss}(\theta_s + v_{ss}w_s) - w_s\frac{\partial}{\partial s}\Big[D_{\eta}(v_s v_{ss} + w_{s}w_{ss})\Big]  \nonumber \\
&+ \Big[\Big(D_\eta - D_\zeta\Big)\frac{\partial}{\partial s}\Big(\theta v_{ss} + \theta^2 w_{ss}\Big)\Big] - \jmath_\xi v_{st}(\theta_t + v_{st} w_s) + w_s\frac{\partial}{\partial t}(\jmath_\eta w_s w_{st} + \jmath_\zeta v_s v_{st}) \nonumber \\
 &- \frac{\partial}{\partial t}\Big[(\jmath_\eta - \jmath_\zeta)(\theta v_{st} + \theta^2 w_{st}) - \jmath_\eta w_{st}\Big] - \frac{w_s}{2}\int_L^s \mathrm{d}s' \, m \int_0^{s'} \mathrm{d}s'' \, \frac{\partial^2}{\partial t^2}\Big(v_s^2 + w_s^2\Big) \Bigg\}, \\
 \jmath_\xi \frac{\partial }{\partial t} &\Big(\theta_{t} + v_{st} w_s\Big) - \frac{\partial }{\partial s}\Big[D_\xi(\theta_s + v_{ss}w_{s})\Big] + (D_\eta - D_\zeta)\Big[(v_{ss}^2 - w_{ss}^2) \theta - v_{ss}w_{ss}\Big]
\nonumber \\
  &- (\jmath_\eta - \jmath_\zeta) \Big[(v_{st}^2 - w_{st}^2)\theta - v_{st} w_{st} \Big] = Q_\theta.
\end{align} 
Here, the subscripts $s$ and $t$ denote differentiating to time and position respectively. $\xi$, $\eta$ and $\zeta$ represent the principle axes of the beam's cross section. $\gamma$ and $Q_\theta$ represents the damping, $D_{\eta,\zeta}$ are the flexural stiffnesses of the beam and $D_{\xi}$ is the torsional stiffness. The moments of inertia are given by $\jmath_{\eta,\zeta,\xi}$. The driving force is $q$ with driving frequency $\Omega$. \\
Now, we consider only vibrations in one direction, so the Eq.A.2 and all terms with $w$ in Eqs.A.1 and A.3 are disregarded. For the torsional mode, the nonlinear effect of torsion warping is taken into account~\cite{Sapountzakis2010:p1853}, where we assume that the ends of the beam are axially immovable. The coupled differential equations (in non-dimensional form, notation from Ref.~\cite{CrespodaSilva:1994p9050}) are now given by:
\begin{align}
v_{tt} &+ \gamma^v v_t + \beta_y v_{ssss} = \frac{\partial }{\partial s} \Bigg\{- v_s \frac{\partial }{\partial s} \Big(\beta_y v_s v_{ss}\Big) -  \frac{\partial }{\partial s}\Big[(1 - \beta_y )\theta^2 v_{ss}\Big] + \jmath_\zeta v_{s}  \frac{\partial }{\partial t}\Big( v_s v_{st}\Big) \nonumber \\
 & -  \frac{\partial }{\partial t} \Big[\Big(\jmath_\zeta - \jmath_\eta\Big)\theta^2 v_{st} + \jmath_\zeta v_{st}\Big] - \frac{v_s}{2}\int_1^s \mathrm{d}s' \, m \int_0^{s'}  \mathrm{d}s'' \,  \frac{\partial^2 }{\partial t^2} \Big(v_s^2 \Big) \Bigg\} + q \cos(\Omega t) , \\
\jmath_\xi \theta_{tt} &- \jmath_{\xi} \gamma^\theta \theta_t + \Big(\beta_\theta + \frac{I_t}{A D_\eta L^2} \tilde{N}\Big)\theta_{ss} - \frac{\rho C_s D_\eta}{\sqrt{m}L^4}  \theta_{sstt} + \frac{E C_s}{L^4} \theta_{ssss}  - \beta_z \theta_s^2 \theta_{ss} \nonumber \\
&- (1 - \beta_y) v_{ss}^2 \theta + (\jmath_\eta - \jmath_\zeta) v_{st}^2 \theta = 0.
\label{eq2}
\end{align}
$\beta_y$ and $\beta_\theta$ are the ratios between two stiffnesses ($\beta_y = D_\zeta / D_\eta$ and $\beta_\theta = D_\xi / D_\eta$) and $A$ is the cross sectional area. The torsion nonlinearity is written as $\beta_z = \frac{3}{2 L^3}E I_n$. The torsion constant $I_t$, the warping constant $C_S$ and the time-dependent tensile axial load $\tilde{N}$ are given by:
\begin{align}
I_t &= \int_S \Big( y^2 + z^2 _ y \frac{\partial \phi_S^P}{\partial z} - z  \frac{\partial \phi_S^P}{\partial y} \Big) \mathrm{d}S,
\nonumber \\
C_S &= \int_S (\phi_S^P)^2 \mathrm{d}S,
\nonumber \\
\tilde{N} &= \frac{1}{2}\frac{EI_p}{l}\int_0^l (\theta_x')^2 \mathrm{d}x.
\end{align}
Here, $S$ is the solid angle and $\phi_S^P$ is the primary warping function. A more detailed description of the nonlinearity in the torsional mode is found in Ref.~\cite{Sapountzakis2010:p1853}. \\ 
\indent \indent To demonstrate the origin of the nonlinear interactions observed in the main text, we now simplify the coupled equations Eq.A.4 and A.5 by applying the Galerkin procedure. The solutions are then written as a linear combination of the linear mode shapes of the cantilever with coefficients, which correspond to the time-dependent vibration, $v = \sum_i F^v_i (s) v_i(t)$ and $\theta = \sum_i F^\theta_i \theta_i(t)$, where $i$ represents the mode number. The mode shapes of the flexural and torsional modes of the cantilever will be discussed. Introducing the operator $\mathcal{L}$, the linear part of Eq.~\ref{eq2} is written as: 
\begin{equation}
\mathcal{L}[F^v] = \beta_y \frac{\partial^4 F^v}{\partial t^4} + \jmath_\zeta \omega^2 F^v = \omega^2 F^v 
\end{equation}
The resonance frequency is denoted as $\omega$. The eigenfunctions can be calculated together with the boundary conditions for a single-clamped cantilever $F^v(0) = F^v_s = F^v_ss(1) = F^v_{sss} = 0$ as:
\begin{align}
F &= [ \cosh(k_1 s) - \cos(k_2 s) - K (\sinh(k_1 s) - k_1/k_2 \sin(k_2 s) ] ,
\nonumber \\
K &= \frac{k_1^2 \cosh(k_1) + k_2^2 \cos(k_2)}{k_1^1 \sinh(k_1) + k_1 k_2 \sin(k_2)},
\end{align} 
The values of $k_{1,2}$ are given by:
\begin{equation}
k_{1,2} = \sqrt{\mp \frac{\jmath_\zeta \omega_B^2}{2 \beta_y} + \sqrt{\Bigg(\frac{\jmath_\zeta \omega_B^2}{2 \beta_y}\Bigg)^2 + \frac{\omega_B^2}{\beta_y} }}.
\end{equation}
The values of $k_1$ and $k_2$ depend on the mode number $i$ and can be calculated via the generating function:
\begin{equation}
k_1^4 + k_2^4 + 2 k_1^2 k_2^2 \cosh(k_1)\cos(k_2) + k_1k_2(k_2 - k_1)^2 \sinh(k_1)sin(k_2) = 0 
\end{equation}
The dimensional resonance frequency of the flexural mode is given by $\omega_{B,i} = \kappa_i (h/L^2) \sqrt{D_\zeta/\jmath_\zeta}$, where $\kappa_i$ is 1.875, 4.695 and 7.855 for $i=1$, 2 and 3. The beam shape of the first two flexural modes are shown in the inset of Fig.~\ref{fig1} of the main text. \\
\indent \indent The torsional mode shapes can be calculated by considering the operator $\mathcal{M}$ with eigenvalues $\omega_T$
\begin{equation}
\mathcal{M}[G] = \frac{\beta_\xi}{\jmath_\xi} \frac{\partial^2 F^\theta}{\partial t^2}  = \omega_T^2 G,
\label{torsion}
\end{equation}
and the corresponding boundary conditions of $F^\theta(0) = F^\theta_s(1) = 0$. Inserting the boundary conditions in Eq.~\ref{torsion}, gives the equation for the torsional mode shapes: $G = \sin[(2i-1)\pi/2  s]$. The resonance frequency of the torsion mode is given by $\omega_T = (2i-1)(\pi/2)\sqrt{\beta_\xi/\jmath_\xi}$. \\
\indent \indent The Galerkin procedure is applied to Eq.A.4 and A.5: i.e. the solutions are written as a linear combination of the eigenmodes. We assume that the flexural mode is only excited around the resonance frequency, accumulating in the equations:
\begin{align}
v^i_{tt} &+ \gamma^v v^i_t + \omega_F^2 v^i = \sum_j \sum_k \sum_l \Big( v^j \Big[\alpha_{1,ijkl} \theta^k \theta^l + \alpha_{2,ijkl} v^k v^l +\alpha_{3,ijkl}  (v^k v^l)_{tt} \Big]  \Big) + f^i \cos(\Omega t),
\nonumber \\
\theta^i_{tt} &+ \gamma^\theta + \omega_T^2 \theta^i =  \sum_j \sum_k \sum_l \Big( \theta^j \Big[\alpha_{4,ijkl} v^k v^l + \alpha_{5,ijkl} \theta^k \theta^l  \Big] \Big).
\end{align}
The above equations show that the nonlinearity is the origin of the modal interactions. Note that for $j=k=l$, the nonlinear equation describing one mode of the cantilever is found. A quadratic coupling is present between two different vibrational modes (for example $k=l$) also follows directly from the cubic nonlinearities. This quadratic coupling is clearly observed in the experiments. In Eq. A.12, the terms linear in $\theta$ are assumed to only modify the resonance frequency $\omega_T$. The coupling (Galerkin) coefficients $\alpha$ are given by the following equations: 
\begin{align}
\alpha_{1,ijkl} &= -(1-\beta_y) \int_0^1 F^i(F^j_{ss} G^k G^l)_{ss} \mathrm{d}s ,
\nonumber \\
\alpha_{2,ijkl} &= -\beta_y \int_0^1 F^i[F^j_s(F^k_s F^l_{ss})_{s}]_s \mathrm{d}s ,
\nonumber \\
\alpha_{3,ijkl} &= -\frac{1}{2} \int_0^1 F^i \Big( F^j_s \int_1^{s''} \int _0^{s'} F^k_s F^l_s \mathrm{d}s \mathrm{d}s'  \Big)_{s''} \mathrm{d}s'' ,
\nonumber \\
\alpha_{4,ijkl} &= \frac{-(1-\beta_y)}{\jmath_\xi}\int_0^1 G^i G^j (F^k F^l)_{ss} \mathrm{d}s,
\nonumber \\
\alpha_{5,ijkl} &= \beta_z \int_0^1G^i (G^j_s G^k_s) G^l_{ss} \mathrm{d}s.
\end{align}
Considering the interactions only between the lowest two modes of the torsional and the flexural mode, the values of the integrals in the coefficients $\alpha$ are given in Table A.1. 
\begin{center}
\begin{table}
\begin{tabular}{c|c|c|c|c|c}
$ijkl$ & $\alpha_1$ & $\alpha_2$ & $\alpha_3$ & $\alpha_4$ & $\alpha_5$ \\
\hline
1111 & 3.213440553 & 40.44066328  & 4.596772482 & 14.71996258  &  -0.01963728194\\
2222 & 317.7598007  & 13418.09334 & 144.7254988 &  45.80067683  & -12.21612177 \\
1211 & -25.80780977 & -102.3196141 & -3.595970428 & -2.107211790 & -0.05545673259 \\
1121 & 5.199336697 & 65.86205943 & -3.595970415  & -45.97052884 &-0.05545673259 \\
1122 & 10.83931447 & 172.7377892 & 25.17415228  & 74.66573509 &  -1.468408253\\
1221 & -20.39723182 & 228.0179031 & 6.117366163 & 22.42570986  & -0.1567290224\\
1212 & -20.39723182 & 2083.845719  &6.117366163 & 22.42570986  & -1.468408253\\
2111 & -25.80780333 & -102.3196141 & -3.595970417 & -2.107211790 & -0.05776014140\\
2211 & 395.6571526 & 172.7376727 & 25.17415228  & 11.08062923  &  -0.1631564726\\ 
2121 & -20.39724250 & 228.0179002 & 6.117366149 & 22.42570986  & -0.1631564726\\
2112 & -20.39724250 & 2083.845647 & 6.117366149 & 22.42570986 & -1.529151963 
\end{tabular}
\label{tab1}
\caption{The values of the integrals in the coefficients $\alpha$ for the interactions between the first and second flexural and torsional modes.}
\end{table}
\end{center}
\indent \indent To give an example, we work out Eq. A.12 for the fundamental and second flexural mode of a cantilever. This case is studied in the experiment and described in Section 3.1. We denote the amplitudes with $v^1=a$ and $v^2=b$) and the coupled equations are given by:
\begin{align}
a_{tt} &+ \gamma_1 a_t + \omega_F^2 a =  \alpha_{2,1111} a^3  + \alpha_{2,1222} b^3 +  (\alpha_{2,1112} + \alpha_{2,1121} + \alpha_{2,1211})a^2 b  
\nonumber \\
&+ (\alpha_{2,1212} + \alpha_{2,1221} + \alpha_{2,1122})  b^2 a + \alpha_{3,1111}  2a(a^2)_{tt}  + \alpha_{3,1222} 2b(b^2)_{tt} + (\alpha_{3,1112} + \alpha_{3,1121}) a(ab)_{tt}
\nonumber \\
 &+ (\alpha_{3,1221} + \alpha{_3,1212}) b(ab)_{tt} + \alpha_{3,1211}  b(a^2)_{tt} + \alpha_{3,1122}  a(b^2)_{tt}   + f_1 \cos(\Omega t), \\
b_{tt} &+ \gamma_2 b_t + \omega_F^2 b =  \alpha_{2,2111} a^3  + \alpha_{2,2222} b^3 +  (\alpha_{2,2112} + \alpha_{2,2121} + \alpha_{2,2211})a^2 b  
\nonumber \\
&+ (\alpha_{2,2212} + \alpha_{2,2221} + \alpha_{2,2122})  b^2 a + \alpha_{3,2111}  2a(a^2)_{tt}  + \alpha_{3,2222} 2b(b^2)_{tt} + (\alpha_{3,2112} + \alpha_{3,2121}) a(ab)_{tt}
\nonumber \\
 &+ (\alpha_{3,2221} + \alpha_{3,2212}) b(ab)_{tt} + \alpha_{3,2211}  b(a^2)_{tt} + \alpha_{3,2122}  a(b^2)_{tt}   + f_2 \cos(\Omega t).
\label{flexflex}
\end{align}


\begin{thebibliography}{25}
\bibitem[{UMf()}]{UMfootnote}
\bibinfo{note}{Between the experiments, the resonance frequency was shifted due
  to thermal drift.}
\end{thebibliography}
\end{document}